\documentstyle[eqsecnum,aps]{revtex}

\begin{document}

\draft

\title{Open universes and avoidance of the cosmological singularity}

\author{ Israel Quiros\thanks{israel@uclv.etecsa.cu}, Rolando Cardenas and Rolando Bonal}
\address{ Departamento de Fisica. Universidad Central de Las Villas. Santa Clara. CP: 54830 Villa Clara. Cuba }

\date{\today}

\maketitle

\begin{abstract}

In reference \cite{iq} a point of view was presented that claims physical equivalence among canonical Einstein's formulation of general relativity and its conformal formulation. It was shown there that, flat (barotropic) Friedmann-Robertson-Walker universes, when modeled with the help of this last formulation of general relativity, are free of the cosmological singularity, in the region $-\frac{3}{2}\leq\omega\leq -\frac{4}{3}$, $0<\gamma<2$ of the parameter space. In the present paper we give conclusive arguments pointing at physical equivalence among conformally related metrics, so this result acquires a new dimension. Based on the argument that any consistent effective theory of spacetime must be invariant under the one-parameter group of transformations of the units of length, time and reciprocal mass, it is shown, also, that canonical general relativity is not such a consistent theory. Conformal general relativity provides a consistent formulation of the laws of gravity instead. We further extend the results of \cite{iq} to open universes by studying the Raychaudhuri equation. It is shown that, in conformal general relativity, these universes are singularity free too in the above region of the parameter space. Exact analytic solutions are found for open Friedmann-Robertson-Walker dust-filled and radiation-filled universes for the particular case when $\omega=-\frac{3}{2}$. 

\end{abstract}


\section{Introduction}

In recent papers \cite{iq} we claimed that canonical Einstein's general relativity(GR) with an extra scalar field and its conformal formulation (called 'Jordan frame GR' therein), are different but physically equivalent representations of a same theory. Our claim was based on the following argument. The spacetime coincidences (coordinates) are not affected by a conformal rescaling of the spacetime metric of the kind

\begin{equation}
\hat g_{ab}=\Omega^2 g_{ab},
\end{equation}
where $\Omega^2$ is a smooth, nonvanishing function on the spacetime manifold. Correspondingly, the experimental observations (measurements), being nothing but just verifications of these coincidences, are unchanged too by the transformations (1.1). This means that canonical GR and its conformal are experimentally indistinguishable. However, we are dissatisfied with the exposition of our ideas in reference \cite{iq} since it may lead to misunderstanding of our view-point and, correspondingly, to misinterpretation of the results presented therein.

The main objection against our claim in ref.\cite{iq} can be based on the following argument. In canonical GR the matter fields couple minimally to the metric that determines metrical relations on a Riemann spacetime, say $\bf{\hat g}$. In this case matter particles follow the geodesics of the metric $\bf{\hat g}$ (on Riemann geometry), and their masses are constant over the spacetime manifold, i.e., it is the metric which matter 'feels' or, if convenient, the 'physical' metric. Under the conformal rescaling (1.1), the matter fields become non-minimally coupled to the conformal metric $\bf g$. Hence, matter particles do not follow the geodesics of this last metric. Furthermore, it is not the metric that determines metrical relations on a Riemann manifold. This line of reasoning leads to the following conclusion. Although canonical GR and its conformal may be physically equivalent theories, nevertheless, the physical metric is that which determines metrical relations on a Riemann spacetime. Hence, the result that, in the conformal formulation of general relativity, flat (barotropic) Friedmann-Robertson-Walker (FRW) universes are free of the cosmological singularity (in the region $-\frac{3}{2}\leq\omega\leq -\frac{4}{3}$, $0<\gamma<2$ of the parameter space), is not relevant. In fact, the conformal (singularity-free) metric $\bf g$, is not the physical metric.

In the present paper, we shall show that the above conclusion is wrong. It is a consequence of the long-standing confusion in the understanding of metric theories of gravity, when conformal transformations of the metric are concerned \cite{fgn}. There is a missing detail when these transformations are studied in the literature. In fact, under the conformal rescaling of the metric (1.1), not only the Lagrangian of the theory is mapped into its conformal Lagrangian, but the spacetime geometry itself is mapped too into a conformal geometry. In this last geometry, metrical relations involve both the conformal metric $\bf g$ and the conformal factor $\Omega^2$ generating the transformation (1.1). Hence, in the conformal Lagrangian, the matter fields should 'feel' both the metric $\bf g$ and the scalar function $\Omega$. I.e., the matter particles would not follow the geodesics of the conformal metric alone. The result is that, under the transformation (1.1), the 'physical' metric of the untransformed geometry is effectively mapped into the 'physical' metric of the conformal geometry. This missing detail is the source of the long-standing confusion in former studies of conformal transformations of the kind (1.1). We aim section {\bf II} of this paper, precisely, at a clarification of this situation, since it is of great importance for the understanding of our view-point.

Yet another question respecting metric theories of spacetime has been long avoided in the literature. It is linked with a program Brans and Dicke outlined in their classical paper \cite{bdk}. In that paper the authors made evident their hope that, the physical content of a given theory of spacetime should be contained in the invariants of the group of position-dependent transformations of units and coordinate transformations. The last part of this program has been already worked-out. By the present, invariance under the group of general coordinate transformations is the minimal requirement any theory, pretending to be a real alternative for the formulation of the laws of gravity, must fulfil\footnotemark\footnotetext{All known metric theories of spacetime, including GR, Brans-Dicke theory and scalar-tensor theories in general, fulfil this requirement.}. It is evident that any consistent formulation of a given effective theory of spacetime must be invariant, also, under the group of the transformations of the units of length, time and mass. However, this part of the program Brans and Dicke outlined in \cite{bdk}, has not been yet worked-out sufficiently. This important subject is treated in detail in section {\bf III} of the present paper. In this section we study the invariance properties of the different formulations of Brans-Dicke (BD) theory and general relativity, under a particular, one-parameter Abelian group of transformations, that can be identified with the group of transformations of the units of length, time and reciprocal mass. We shall show that the only consistent formulation of the laws of gravity (among those studied here), is the conformal representation of general relativity.

In this context, the result presented in ref.\cite{iq} that flat FRW universes, in the frame of conformal GR, are free of the cosmological singularity (in the given region of the parameter space), acquires a new dimension. The occurrence of spacetime singularities may be interpreted as a result of a wrong choice of the formulation of the given spacetime theory. In section {\bf IV} we further extend the result of paper \cite{iq} to open FRW universes.

A discussion on the results obtained in the present and former papers (\cite{iq}), and on the physical significance of our view-point, is given in section {\bf V}.

\section{Effective theories of spacetime and conformal geometries}

In this section we shall investigate the effect of a conformal transformation of the kind (1.1) on the laws of gravity and on the geometry. For illustration we shall study general relativity with an extra scalar field. The canonical Lagrangian for this theory is

\begin{equation}
\hat{\cal L}_{GR}=\sqrt{-\hat g}(\hat R -\alpha(\hat \nabla \hat \phi)^2)+16\pi\sqrt{-\hat g} L_{matter},
\end{equation}
where $\hat R$ is the Ricci scalar given in terms of the untransformed metric $\bf{\hat g}$, $(\hat\nabla\hat\phi)^2=\hat g^{mn}\hat\phi_{,m}\hat\phi_{,n}$, $\alpha$ is a free parameter ($\alpha\geq 0$) and $L_{matter}$ is the Lagrangian of the matter fields. When $\phi$ is a constant, or when $\alpha=0$ (arbitrary $\phi$), we recover usual Einstein's theory. Under the conformal rescaling (1.1) with $\Omega^2=e^{\hat\phi}$, the Lagrangian (2.1) is mapped into its conformal Lagrangian

\begin{equation}
{\cal L}_{GR}=\sqrt{-g}e^{\hat\phi}(R -(\alpha-\frac{3}{2})(\nabla \hat \phi)^2)+16\pi\sqrt{-g}e^{2\hat\phi} L_{matter},
\end{equation}
where now $R$ is the Ricci scalar given in terms of the conformal metric $\bf g$. This Lagrangian can be given a more usual, Brans-Dicke form, after the change of variable $\hat\phi\rightarrow\phi=e^{\hat\phi}$,

\begin{equation}
{\cal L}_{GR}=\sqrt{-g}(\phi R -(\alpha-\frac{3}{2})\frac{(\nabla\phi)^2}{\phi})+16\pi\sqrt{-g}\phi^2 L_{matter}.
\end{equation}

Due to the minimal coupling of the scalar field $\hat\phi$ to the curvature in canonical GR (Lagrangian (2.1)), the effective gravitational constant $\hat G$ (set equal to unity in (2.1)) is a real constant. The minimal coupling of the matter fields to the metric in (2.1), warrants that matter particles follow the geodesics of the metric $\bf{\hat g}$, that is the metric which defines metrical relations on Riemann spacetimes. Hence, the inertial mass $\hat m$ of some elementary particle is constant too over the spacetime. This leads that the dimensionless gravitational coupling constant $\hat G \hat m^2$ ($c=\hbar=1$) is constant in spacetime unlike BD theory, where $\hat G \hat m^2$ evolves like $\phi^{-1}$. This is a conformal invariant feature of general relativity since dimensionless constants do not change under (1.1). In other words, in conformal general relativity, the dimensionless gravitational constant $Gm^2$ is a constant as well ($G$ and $m$ are the effective gravitational constant and the inertial mass of some elementary particle respectively, given in the conformal GR). However, in this last case (Lagrangian (2.2) or (2.3)), the effective gravitational constant varies like $G\sim e^{-\hat\phi}$ (or $\sim \phi^{-1}$). Since $\hat m^2=Gm^2$ ($\hat G=1$), then, in this formulation of general relativity, particle masses vary over the spacetime manifold like

\begin{equation}
m=e^{\frac{1}{2}\hat\phi}\hat m.
\end{equation}

According to Dicke \cite{dk}, the conformal transformation (1.1) (in \cite{dk} $\Omega^2=e^{\hat\phi}=\phi$) can be interpreted as a transformation of the units of length, time and reciprocal mass. In fact, if one chooses the arc-length as one's unit of length and time, and the inertial mass of some elementary particle as one's unit of mass, then, we see that $ds=e^{-\frac{1}{2}\hat\phi}d\hat s$, while $m^{-1}=e^{-\frac{1}{2}\hat\phi}\hat m^{-1}$, i.e., these measuring scales change in the same way over the manifold.

A careful looking at eqs.(2.1), (2.2,2.3) shows that the Einstein's laws of gravity derivable from (2.1), change under the units transformation (1.1). This seems to be a serious drawback of canonical GR (and BD theory and scalar-tensor theories in general) since, it is obvious that, in any consistent theory of spacetime, the laws of physics must be invariant under a change of the units of length, time and mass. This subject will be discussed in detail in the next section. It will be shown that the transformation (1.1) with $\Omega^2=e^{\hat\phi}=\phi$, can not be taken properly as a units transformation. It is just a transformation that allows 'jumping' from one formulation of the given spacetime theory into its conformal.

In ref.\cite{iq} we claimed that canonical GR (Lagrangian (2.1)) and its conformal (Lagrangian (2.2) or (2.3)) are physically equivalent theories, since they are indistinguishable from the observational point of view. However, it is of common belief, that only one of the conformally related metrics is the 'physical' metric, i.e., that which determines metrical relations on the spacetime manifold. The reasoning line leading to this conclusion is based on the following analysis. Take, for instance, general relativity with an extra scalar field. Due to the minimal coupling of the matter fields to the metric in (2.1), the matter particles follow the geodesics of the metric $\bf{\hat g}$,

\begin{equation}
\frac{d^2 x^a}{d\hat s^2}+\hat\Gamma^a_{mn}\frac{dx^m}{d\hat s}\frac{dx^n}{d\hat s}=0,
\end{equation}
where $\hat\Gamma^a_{bc}=\frac{1}{2}\hat g^{an}(\hat g_{bn,c}+\hat g_{cn,b}-\hat g_{bc,n})$ are the Christoffel symbols of the metric $\bf{\hat g}$. These coincide with the geodesics of the Riemann geometry, where metrical relations are given by $\bf{\hat g}$ through the expressions for the scalar product of two vectors $\bf{\hat X}$ and $\bf{\hat Y}$, $\hat g({\bf\hat X},{\bf\hat Y})=\hat g_{mn}\hat X^m \hat Y^n$, the line-element $d\hat s^2=\hat g_{mn} dx^m dx^n$, etc. It is the reason why canonical GR, based on the Lagrangian (2.1), is naturally linked with Riemann geometry\footnotemark\footnotetext{The same is true for the Jordan frame formulation of Brans-Dicke theory since, in this frame, the matter fields couple minimally to the spacetime metric.}. The units of this geometry are constant over the manifold. This requirement is realized in canonical general relativity and in the Jordan frame formulation of Brans-Dicke theory, through the assumption that there exists a large class of physical systems (such like atoms) having properties that are independent of location \cite{bdk}. This is equivalent to say that one can take some quantities associated with these systems (for instance the atom radius and transition energies) as one's units of measurement. These will be constant over the manifold. Hence, for instance, the arc-length between two successive events on a geodesic curve will be point-independent, as required by the behavior of the units of measure of Riemann geometry\cite{haw}.

On the other hand, since the matter fields are non-minimally coupled to the metric in the conformal GR, the matter particles would not follow the geodesics of the conformal metric $\bf g$. These will follow curves that are solutions of the equation conformal to (2.5) instead

\begin{equation}
\frac{d^2 x^a}{ds^2}+\Gamma^a_{mn}\frac{dx^m}{ds}\frac{dx^n}{ds}+\frac{\phi_{,n}}{2\phi}(\frac{dx^n}{ds}\frac{dx^a}{ds}-g^{na})=0,
\end{equation}
where now $\Gamma^a_{bc}$ are the Christoffel symbols of the metric $\bf g$ conformal to ${\bf\hat g}$. Hence, if one assumes that the spacetime geometry is fixed to be Riemannian and that it is unchanged under the conformal rescaling (1.1) with $\Omega^2=\phi$, one effectively arrives at the conclusion that ${\bf\hat g}$ is the 'physical' metric. However, this assumption is wrong and is the source of the long-standing confusion in the understanding of the meaning of the conformal transformations of the metric (1.1)\cite{fgn}. We shall show this in the remainder of this section.

\subsection{Conformal Riemann geometry}
 
Let $\lambda(t)$ be a curve on the spacetime manifold with local coordinates $x^a(t)$ and let $\bf X$ with local coordinates $X^a=\frac{dx^a}{dt}$, be a vector tangent to $\lambda(t)$. The covariant derivative of a given vector field ${\bf\hat Y}$ along $\lambda$ is given by\footnotemark\footnotetext{We use the symbology of reference \cite{haw}.}

\begin{equation}
\frac{\hat D\hat Y^a}{\partial t}=\frac{\partial\hat Y^a}{\partial t}+\hat\gamma^a_{mn}\hat Y^m\frac{dx^n}{dt},
\end{equation}
where $\hat\gamma^a_{bc}$ is a symmetric connection on the manifold. The vector ${\bf\hat Y}$ is said to be parallely transported along $\lambda$ if $\frac{\hat D{\bf\hat Y}}{\partial t}=0$. In particular, if one considers the covariant derivative of the tangent vector itself along $\lambda$, then, the curve $\lambda(t)$ is said to be a geodesic curve if $\frac{\hat D}{\partial t}(\frac{\partial}{\partial t})_{\lambda}$ is parallel to $(\frac{\partial}{\partial t})_{\lambda}$\cite{haw}. In other words, there exists a function $f$ such that $X^n \hat D_n X^a=f X^a$. For such a curve, one can find an affine parameter $\hat v(t)$ along the curve such that $\frac{\hat D}{\partial\hat v}(\frac{\partial}{\partial\hat v})_{\lambda}=0$. The associated tangent vector ${\bf\hat V}=(\frac{\partial}{\partial\hat v})_{\lambda}$ is parallel to ${\bf\hat X}$ and its scale is determined by $\hat V(\hat v)=1$. It obeys the equations

\begin{equation}
\hat V^n \hat D_n \hat V^a=0.
\end{equation}

Given a metric ${\bf\hat g}$ on the manifold $\hat{\cal M}$, the Riemann geometry is fixed by the following postulate. There is a unique torsion-free (symmetric) connection on $\hat{\cal M}$ defined by the condition that the covariant derivative of ${\bf\hat g}$ is zero, i.e. $\hat D_c \hat g_{ab}=0$. With the connection defined in such a way, parallel transfer of vectors, for instance, ${\bf\hat Y}$

\begin{equation}
\frac{\hat D\hat Y^a}{\partial t}=\frac{\partial\hat Y^a}{\partial t}+\hat\gamma^a_{mn}\hat Y^m\frac{dx^n}{dt}=0,
\end{equation}
preserves scalar products defined by ${\bf\hat g}$, in particular

\begin{equation}
dg({\bf\hat Y},{\bf\hat Y})=0,
\end{equation}
where $g({\bf\hat Y},{\bf\hat Y})=\hat g_{mn}\hat Y^m \hat Y^n$. The laws of parallel transport (2.9) and length preservation (2.10) together lead that the symmetric connection $\hat\gamma^a_{bc}$ on a Riemann manifold, coincides with the Christoffel symbols of the metric ${\bf\hat g}$; $\hat\gamma^a_{bc}=\hat\Gamma^a_{bc}$. Then, eq.(2.8) defining a geodesic curve on the Riemann manifold, can be written as

\begin{equation}
\frac{d^2 x^a}{d\hat v^2}+\hat\Gamma^a_{mn}\frac{dx^m}{d\hat v}\frac{dx^n}{d\hat v}=0.
\end{equation}

If ${\bf\hat V}$ is a time-like vector, then, in particular, the affine parameter $\hat v$ can be set equal to the arc-length $\hat s$.

Suppose that, under the conformal rescaling (1.1) with $\Omega^2=\phi$, the vectors ${\bf\hat Y}$ and ${\bf\hat V}$ transform in the following ways,

\begin{equation}
\hat Y^a=h(\phi)Y^a,\;\; \hat V^a=j(\phi)V^a,
\end{equation}
where $h$ and $j$ are smooth functions of the scalar field $\phi$. Hence, the law (2.10) of preservation of length of a vector ${\bf\hat Y}$ in Riemann geometry, is transformed into the following law of length transport for the conformal vector ${\bf Y}$,

\begin{equation}
dg({\bf Y},{\bf Y})=-d[\ln(\phi h^2)]g({\bf Y},{\bf Y}).
\end{equation}

This resembles the law of length transport in Weyl geometry. Hence, given a Riemann geometry on $\hat{\cal M}$, under (1.1) with $\Omega^2=\phi$, it is transformed into a Weyl geometry on $\cal M$. The parallel transport law conformal to (2.9), can be written as

\begin{equation}
\frac{D Y^a}{\partial t}+\frac{\partial}{\partial t}(\ln\; h)\; Y^a=0,
\end{equation}
where the following definition for the conformal covariant derivative has been used

\begin{equation}
\frac{D Y^a}{\partial t}=\frac{\partial Y^a}{\partial t}+\gamma^a_{mn} Y^m\frac{dx^n}{dt}=0.
\end{equation}

$\gamma^a_{bc}$ is the symmetric connection on the Weyl manifold. It is related with the Christoffel symbols of the conformal metric ${\bf g}$ through

\begin{equation}
\gamma^a_{bc}=\Gamma^a_{bc}+\frac{1}{2}\phi^{-1}(\phi_{,b}\delta^a_c+\phi_{,c}\delta^a_b-g_{bc} g^{an}\phi_{,n}).
\end{equation}

If one applies eq.(2.14) to the vector ${\bf V}=(\frac{\partial}{\partial v})_{\lambda}$ that is tangent to the curve $\lambda$ (it fulfils $V^n D_n V^a=f\;\; V^a$, where$f=-\frac{\partial}{\partial v}(\ln\; j)$), then, we obtain the local coordinate expression for the geodesic equation that is conformal to (2.11),

\begin{equation}
\frac{d^2 x^a}{dv^2}+\Gamma^a_{mn}\frac{dx^m}{dv}\frac{dx^n}{dv}+\frac{\phi_{,n}}{2\phi}(2\frac{dx^n}{dv}\frac{dx^a}{dv}-g^{na})+\frac{j_{,n}}{j}\frac{dx^n}{dv}\frac{dx^a}{dv}=0.
\end{equation}

The main feature of Weyl geometry, being based on the laws of parallel transport (2.14) and length transport (2.13), is that the units of measure of this geometry are point-dependent. In particular, the law (2.13) applied to the arc-length, leads that the line-element changes from point to point in spacetime like $ds^2\sim\phi^{-1}$. In other words, the arc-length between two neighboring events on a geodesic curve is different for different points on the curve. Hence, Weyl geometry represents a generalization of Riemann geometry to include units of measure with point-dependent length.

\subsection{Linkage between effective theories of spacetime and conformal geometries}

When an effective theory of spacetime is approached, some conclusions raise. First, theories with minimal coupling of the matter fields to the metric are naturally linked with Riemann geometry with constant units of measure, while theories with non-minimal coupling are linked with Weyl geometry with varying length units of measure. Conformal GR is an example of this last case. In fact, we see that the geodesic equation for a time-like vector ${\bf V}=(\frac{\partial}{\partial s})_{\lambda}$ on Weyl geometry (eq.(2.17) with $v=s$ and $j=\phi^{-\frac{1}{2}}$) exactly coincides with the equation (2.6) defining free-motion time-like path in conformal general relativity. The second conclusion is connected with the fact that, in Weyl geometry, metrical relations are given by the metric $\bf g$ and by the scalar field $\phi$. This last field determines how metrical relations change from point to point on the Weyl manifold (see eq.(2.13)). Hence, on Weyl geometry, any matter field would 'feel' both the metric $\bf g$ and the scalar field $\phi$, i.e., matter would be coupled both to $\bf g$ and to $\phi$ (non-minimal coupling). Following the same reasoning-line leading to the identification of ${\bf\hat g}$ as the physical metric in canonical general relativity, we reach to the following crucial conclusion. The conformal transformation (1.1) with $\Omega^2=\phi$, maps the 'physical' metric ${\bf\hat g}$ on Riemann geometry into the 'physical' metric $\bf g$ on its conformal (Weyl) geometry. Correspondingly, geodesic curves on Riemann spacetimes are mapped into geodesic curves on Weyl spacetimes. In particular, incomplete geodesics on a Riemann spacetime can, in principle, be mapped into complete geodesics on the conformal spacetimes. Hence, spacetime singularities that plage the canonical (Riemannian) general relativity, may be avoided in conformal general relativity linked with Weyl geometry. This subject will be treated in section {\bf IV}, for Friedmann-Robertson-Walker spacetimes.

\section{Effective theories of spacetime and transformations of units}

We shall study two kinds of Lagrangians for pure gravity,

\begin{equation}
{\cal L}_1=\sqrt{-g}(R-\alpha(\nabla\phi)^2),
\end{equation}
and

\begin{equation}
{\cal L}_2=\sqrt{-g}(\phi R-(\alpha-\frac{3}{2})\frac{(\nabla\phi)^2}{\phi}),
\end{equation}
in respect to their transformation properties under rescalings of the units of length, time and reciprocal mass of the kind (1.1)\footnotemark\footnotetext{Lagrangian (3.2) can be obtained from (3.1) if we rescale ${\bf g}\rightarrow\phi{\bf g}$ and change $\phi\rightarrow\ln\phi$.}. We shall interested, in particular, in the following conformal transformation:

\begin{equation}
\tilde g_{ab}=\phi^\sigma\; g_{ab},
\end{equation}
where $\sigma$ is some arbitrary parameter. Under (3.3) the Lagrangian ${\cal L}_1$ changes into

\begin{equation}
\tilde{{\cal L}}_1=\sqrt{-\tilde g}(\phi^{\sigma}\tilde R+((3\sigma-\frac{3}{2}\sigma^2)\phi^{-2-\sigma}-\alpha\; \phi^\sigma)(\tilde\nabla\phi)^2),
\end{equation}
so the laws of gravity it describes, change under (3.3). In particular, in the conformal frame (magnitudes with tilde), the effective gravitational constant depends on $\phi$ due to the non-minimal coupling between the scalar field $\phi$ and the curvature. Lagrangian (3.2), on the other hand, is mapped into

\begin{equation}
\tilde{{\cal L}}_2=\sqrt{-\tilde g}(\phi^{1-\sigma}\tilde R-\frac{(\alpha-\frac{3}{2}-3\sigma+\frac{3}{2}\sigma^2)}{(1-\sigma)^2}\phi^{\sigma-1}(\tilde\nabla\phi^{1-\sigma})^2).
\end{equation}

Hence, if we introduce a new scalar field variable

\begin{equation}
\tilde\phi=\phi^{1-\sigma},
\end{equation}
and redefine the free parameter of the theory

\begin{equation}
\tilde\alpha=\frac{\alpha+3\sigma(\sigma-2)}{(1-\sigma)^2},
\end{equation}
then, the Lagrangian $\tilde{{\cal L}}_2$ (eq.(3.5)) can be put in the following form

\begin{equation}
\tilde{{\cal L}}_2=\sqrt{-\tilde g}(\tilde\phi\;\tilde R-(\alpha-\frac{3}{2})\frac{(\tilde\nabla\tilde\phi)^2}{\tilde\phi}),
\end{equation}
i.e., the Lagrangian ${\cal L}_2$ is invariant in form under the conformal transformation (3.3), the scalar field redefinition (3.6) and the parameter transformation (3.7). To our knowledge, these transformations were first studied in \cite{far}. The composition of two successive transformations (3.3), (3.6) and (3.7), with parameters $\sigma_1$ and $\sigma_2$, gives a transformation of the same kind with parameter $\sigma_3$,

\begin{equation}
\sigma_3=\sigma_1+\sigma_2-\; \sigma_1 \; \sigma_2.
\end{equation}

The identity transformation is that with $\sigma=0$, while the inverse of the transformation with parameter $\sigma$, is a transformation with parameter $\bar\sigma=-\frac{\sigma}{1-\sigma}$. Hence, if we exclude the value $\sigma=1$, then (3.3), (3.6) and (3.7) form a one-parameter Abelian group of transformations ($\sigma_3\;(\sigma_1,\sigma_2)=\sigma_3\;(\sigma_1,\sigma_2)$). Since (3.3) can be interpreted as a transformation of the units of length, time and reciprocal mass\cite{dk}, we shall call this as the one-parameter group of transformations of these units. We see that the transformation (3.3) with $\sigma=1$ does not belong to this group (the inverse does not exist) and, hence, it can not be interpreted properly as a units transformation\footnotemark\footnotetext{Transformation of units is a group-theoretic technique\cite{dk}.}.

A very important conclusion raises. Since any consistent theory of spacetime must be invariant under the one-parameter group of transformations of the units of length, time and mass, then, spacetime theories whose Lagrangian for pure gravity is of the kind ${\cal L}_1$, are not consistent theories while, those based on Lagrangians of the kind ${\cal L}_2$ may, in principle, be consistent formulations of a spacetime theory. Hence, for instance, canonical GR and the Einstein frame formulation of BD theory are not consistent formulations of the laws of gravity.

We shall now consider, separately, the following matter Lagrangians

\begin{equation}
\sqrt{-g}\;\phi^2\;L_{matter},
\end{equation}
and

\begin{equation}
\sqrt{-g}\;L_{matter}.
\end{equation}

The Lagrangian (3.11) shows minimal coupling of matter to the metric, while Lagrangian (3.10) shows non-minimal coupling. Under (3.3) the Lagrangian (3.10) is transformed in the following way,

\begin{equation}
\sqrt{-g}\;\phi^2\;L_{matter}=\sqrt{-\tilde g}\;\phi^{2-2\sigma}\; L_{matter},
\end{equation}
and, hence, considering the scalar field redefinition (3.6), we complete the demonstration that (3.10) is invariant in form under the one-parameter group of transformations of the units of length, time and reciprocal mass. Unfortunately, it is straightforward that the Lagrangian (3.11) with minimal coupling, is not invariant under this group of units transformations. The consequence is that Brans-Dicke theory (its Jordan frame formulation) based on the Lagrangian\cite{bdk}

\begin{equation}
{\cal L}_{BD}={\cal L}_2+16\pi\sqrt{-g}\; L_{matter},
\end{equation}

is not yet a consistent theory of spacetime. The only surviving theory is the conformal general relativity based on the Lagrangian (2.3), i.e.,

\begin{equation}
{\cal L}_{GR}={\cal L}_2+16\pi\sqrt{-g}\;\phi^2\; L_{matter}.
\end{equation}

This theory provides a consistent formulation of the laws of gravity. In fact, the laws derived from (2.3) (or (3.14)) are invariant under the one-parameter Abelian group of units transformations (3.3), (3.6) and (3.7) as required. We think, it is a very serious argument to take in mind while studying different alternatives for a final theory of spacetime.

It is not casual that the laws under which Weyl geometry is based (laws of parallel transport (2.14) and length transport (2.13)), are invariant too under the transformations of this group. In fact, as we have already shown, the conformal formulation of general relativity (Lagrangian (3.14)), being invariant under the transformations of this group, is naturally linked with Weyl geometry. Unlike this, Riemann geometry is not invariant under (3.3) and (3.6). The parallel transport law (2.9) and the length preservation law (2.9), change under these transformations. Hence, Riemann geometry is not a consistent framework for the interpretation of the laws of gravity while Weyl geometry does it.

Finally, we shall remark that the conformal rescaling of the metric (1.1) with $\Omega^2=\phi$ (the case $\sigma=1$ in (3.3)) can not be interpreted properly as a units transformation, since it does not belong to the one-parameter Abelian group studied above. The role of this particular transformation is just to warrant 'jumping' from the inconsistent canonical formulation of GR, into the consistent conformal formulation of this theory. At the same time, it permits 'jumping' from the inconsistent Riemann geometry into the consistent Weyl geometry.

\section{Conformal general relativity and the cosmological singularity}
 
In reference \cite{iq} we showed that the cosmological singularity that is always present in the canonical (Einstein frame) formulation of general relativity, may be removed when we work in conformal representation of this theory (in the 'plus' branch of the solution)\cite{iq}. This result was obtained for Friedmann-Robertson-Walker(FRW), flat universes filled with a barotropic perfect fluid. The absence of singularities in the conformal frame was expected since, in the '+' branch of the solution $R_{mn}k^m k^n$ may be negative definite (in the given region of the parameter space) for any non-spacelike vector $\bf k$. This means that the relevant singularity theorems may not hold\footnotemark\footnotetext{The vanishing of the cosmological singularity can be explained, in other words, as due to the fact that, under (1.1) with $\Omega^2=\phi$, incomplete geodesics on a Riemann FRW spacetime, are mapped into complete geodesics on the conformal (Weyl) spacetime, in a given region of the parameter space.}. This is in contradiction with the canonical formulation where $\hat R_{mn} \hat k^m \hat k^n$ is non-negative and a space-like singularity at the beginning of time $t=0$ always occurs. These are not alternative theories, but just alternative representations of the same theory \cite{iq}. Both of them are observationally equivalent. However,  observational evidence should be interpreted either on the grounds of Riemann geometry (with constant units of measure) or on the grounds of Weyl geometry (with varying length units of measure), depending on the formulation of general relativity we chose for modeling the universe. 

In the light of our previous discussion in sections {\bf II} and {\bf III}, the result that the cosmological singularity may be removed (in some region of the parameter space of the theory), in the conformal representation of general relativity, is very interesting since, it suggests that the spacetime singularity may be a spurious entity due to a wrong choice of the formulation of GR. In fact, the conformal transformation (1.1) with $\Omega^2=\phi$, effectively removes the cosmological singularity occurring in canonical Einstein's GR on Riemann manifolds. As we have already shown, this last representation of general relativity provides an inconsistent formulation of the laws of gravity. Weyl geometry, on the contrary, is a consistent framework where to formulate the laws of gravitation. 

Another argument in this direction is connected with the fact that the canonical formulation of general relativity is a classical theory of spacetime and, it is the hope that, when a final quantum theory of gravity will be worked out, the singularity will be removed. In the conformal representation of general relativity no singularity occurs even without including quantum considerations\cite{iq}. Other arguments come from string theory where a scalar field (the dilaton) is always coupled to the curvature\cite{gsw}.

In this section we shall further extend the results of \cite{iq} to open FRW universes filled with a barotropic perfect fluid. 

The field equations derivable from (2.1), for open FRW universes, can be reduced to the following equation for determining the untransformed scale factor $\hat a$,

\begin{equation}
(\frac{\dot{\hat a}}{\hat a})^2-\frac{1}{\hat a^2}=\frac{M}{\hat a^{3\gamma}}+\frac{\alpha\;N}{6\hat a^6},
\end{equation}
where, in order that the results here were consistent with the symbology used in ref.\cite{iq}, we should realize that $\alpha=\omega+\frac{3}{2}$. $N$ and $M$ are arbitrary integration constants and the barotropic index $\gamma$ is in the range $0<\gamma< 2$. While deriving eq.(4.1) we have taken into account that, in the untransformed frame, the ordinary matter energy density is given by $\hat\mu=\frac{3}{8\pi}\frac{M}{\hat a^{3\gamma}}$. After integrating once the wave equation $\hat\Box\hat\phi=0$ for the scalar field $\hat\phi$, we obtain:

\begin{equation}
\dot{\hat\phi}=\pm\frac{\sqrt{N}}{\hat a^3}.
\end{equation}

This equation has been considered also while deriving eq.(4.1). The curvature scalar for an open FRW universe is found to be

\begin{equation}
\hat R=\frac{3M}{\hat a^{6}}[(4-3\gamma)\hat a^{3(2-\gamma)}-\frac{\alpha\;N}{3M}].
\end{equation}

Hence at $\hat a=0$ there is a curvature singularity which corresponds to the initial cosmological singularity. In fact, with the help of eq.(4.1) for $\hat a\ll 1$, we find that $t\sim\hat a^3$. Hence the correspondence $\hat a\rightarrow 0\;\Leftrightarrow\;t\rightarrow 0$.

Our goal here is to show that this curvature singularity is removed when we 'jump' to the conformal formulation of general relativity. For this purpose we shall first study the Raychaudhuri equation for a congruence of fluid lines and then we shall study, in detail, the behaviour of the relevant magnitudes and relationships in the conformal formulation of GR. Finally we shall study the particular case with $\alpha=0$ ($\omega=-\frac{3}{2}$) for dust-filled and radiation-filled universes since, in these cases, exact analytic solutions are easily found.

The Raychaudhuri equation for a congruence of fluid lines without vorticity and shear, with the time-like tangent vector $\hat k^a=\delta^a_0$, can be written (in the canonical formulation of GR) as:

\begin{equation}
\dot{\hat\theta}=-\hat R_{00}-\frac{1}{3}\hat\theta^2,
\end{equation} 
where the overdot means derivative with respect to the untransformed proper time $t$ and $\hat\theta$ is the volume expansion. In eq.(4.4) we took the reversed sense of time $-\infty\leq t\leq 0$, i.e. $\hat a$ runs from infinity to zero. This equation can be finally written as: 

\begin{equation}
\dot{\hat\theta}=-\frac{3}{\hat a^2}-\frac{9\gamma M}{2\hat a^{3\gamma}}-\frac{3}{2}\frac{\alpha\;N}{\hat a^6}.
\end{equation}

From it one sees that all terms in the right-hand side induce contraction and hence a spacetime singularity is expected to occur (the global singularity at $\hat a=0$).

The evolution of the volume expansion in the conformal frame, given in terms of the untransformed scale factor, can be easily found from eq.(4.5) if we realize that $\theta=e^{\frac{\hat\phi}{2}}(\hat\theta-\frac{3}{2}\hat\phi)$. We find that

\begin{equation}
(\frac{d\theta}{d\tau})^\pm=3\frac{e^{\hat\phi^\pm}}{\hat a^6}\{-\hat a^4-\frac{3}{2}\gamma M\hat a^{3(2-\gamma)}-\frac{1}{2}(\alpha+\frac{1}{2})N\pm\frac{\sqrt{N}}{2}\sqrt{\hat a^4+M\hat a^{3(2-\gamma)}+\frac{1}{6}\;\alpha\;N}\},
\end{equation}
where $\tau$ is the proper time in the conformal frame. It is related with $t$ through $d\tau=e^{-\frac{1}{2}\hat\phi^\pm}dt$. The '+' and '-' signs in eq.(4.6) correspond to two possible branches of the solution in conformal general relativity. From (4.6) one sees that for the '+' branch of the solution, the last term in brackets induces expansion.

We are interested now in the limiting case $\hat a\ll 1$, since the singularity in the untransformed frame is found at $\hat a=0$. In this case for $\alpha=0$ ($\omega=-\frac{3}{2}$) eq.(4.6) can be written as:

\begin{equation} 
(\frac{d\theta}{d\tau})^\pm\approx\pm\frac{3\sqrt{NM}e^{\hat\phi_0}}{2\hat a^{\frac{3}{2}(\gamma+2)}}\exp[\mp\frac{2}{3}\sqrt{\frac{N}{M}}\frac{\hat a^{-\frac{3}{2}(2-\gamma)}}{(2-\gamma)}],
\end{equation}
for $\gamma>\frac{2}{3}$. $\hat\phi_0$ is some integration constant. For $\gamma=\frac{2}{3}$ we obtain:

\begin{equation} 
(\frac{d\theta}{d\tau})^\pm\approx\pm\frac{3\sqrt{N(M+1)}e^{\hat\phi_0}}{2\hat a^4}\exp[\mp\frac{1}{2}\sqrt{\frac{N}{M+1}}\hat a^{-2}],
\end{equation}
while for $\gamma<\frac{2}{3}$;

\begin{equation} 
(\frac{d\theta}{d\tau})^\pm\approx\pm\frac{3\sqrt{N}e^{\hat\phi_0}}{2\hat a^4}\exp[\mp\frac{\sqrt{N}}{2}\hat a^{-2}].
\end{equation}

For arbitrary positive $\alpha$ ($\omega>-\frac{3}{2}$), in the limit $\hat a\ll 1$, eq.(4.6) can be written in the following way:

\begin{equation} 
(\frac{d\theta}{d\tau})^\pm\approx\frac{N\;e^{\hat\phi_0}}{2\;\hat a^{6\mp\sqrt{\frac{6}{\alpha}}}}(-3\alpha\pm 2\sqrt{6\alpha}-\frac{3}{2}) ,
\end{equation}
for $0<\gamma<2$.

A careful analysis of eq.(4.6) shows that, for big $\hat a$ the first three terms in brackets prevail over the last one and, consequently, contraction is favored until $\hat a$ becomes sufficiently small ($\hat a\ll 1$). In this case, when $\alpha$ is in the range $0\leq\alpha\leq\frac{1}{6}$ ($-\frac{3}{2}\leq\omega\leq -\frac{4}{3}$), in the '+' branch of the solution, there are not enough conditions for further contraction and the formation of the global singularity is not allowed. A cosmological wormhole\cite{vh} is obtained instead.
For further analysis of what happen in this case, we need to write down the relevant magnitudes and relationships, in the conformal frame, in terms of the untransformed scale factor. We shall interested in the behaviour of these magnitudes and relationships for small $\hat a\ll 1$ (when the condition for further contraction ceases to hold), in the '+' branch of the solution. In this case the scale factor is found to be
  
\begin{equation}
a^+\approx e^{-\frac{1}{2}\hat\phi_0}\hat a^{1-\frac{1}{2}\sqrt{\frac{6}{\alpha}}}.
\end{equation}

The conformal Ricci scalar is

\begin{equation}
R^+\approx\frac{3}{2}N e^{\hat\phi_0}\hat a^{\sqrt{\frac{6}{\alpha}}-6},
\end{equation}
while for the proper time $\tau$ we have that 

\begin{equation}
\tau^+\approx\frac{2 e^{-\frac{1}{2}\hat\phi_0} \hat a^{3-\frac{1}{2}\sqrt{\frac{6}{\alpha}}}}{\sqrt{\frac{\alpha\;N}{6}}(6-\sqrt{\frac{6}{\alpha}})},
\end{equation}
for $\alpha\neq\frac{1}{6}$ and

\begin{equation}
\tau^+\approx\frac{e^{-\frac{1}{2}\hat\phi_0}}{\sqrt{\frac{\alpha\;N}{6}}}\ln \hat a,
\end{equation}
for $\alpha=\frac{1}{6}$. Hence, if we choose the '+' branch of the solution, for $\alpha\geq\frac{1}{6}$, $R^+$ is bounded for $\hat a\rightarrow 0$. In this limit $a^+\rightarrow +\infty$ while $\tau^+\rightarrow -\infty$. A similar analysis shows that, for big $\hat a$, $\hat a\rightarrow\infty\Rightarrow a^\pm\rightarrow\infty$ and $\tau^\pm\rightarrow +\infty$. For intermediate values in the range $0<\hat a<\infty$ the curvature scalar $R^+$ is well behaved and bounded. The scale factor $a$ is a minimum at some intermediate time $\tau_*$. Hence in the conformal representation of general relativity, if we choose the '+' branch of the solution, the following picture takes place. If we restrict $\omega$ to fit into the range $-\frac{3}{2}\leq\omega\leq -\frac{4}{3}$ and for $0<\gamma<2$, then, the universe evolves from the infinite past $\tau=-\infty$ when he had an infinite size, through a bounce at some intermediate $\tau_*$ when he reached a minimum size $a_*$, into the infinite future $\tau=+\infty$ when he will reach again an infinite size. As illustrations to this behaviour we shall study the particular cases with $\omega=-\frac{3}{2}$ ($\alpha=0$) for dust-filled and radiation-filled universes since, in these very particular situations exact analytic solutions can be easily found.

For a radiation-filled universe ($\gamma=\frac{4}{3}$) the equation (4.1) with $\alpha=0$ can be written as:

\begin{equation}
\dot{\hat a}=\sqrt{M\hat a^{-2}+1},
\end{equation}
and, after integration we obtain for the untransformed scale factor

\begin{equation}
\hat a=\sqrt{t^2-M}.
\end{equation}

The proper time $t$ is constrained to the range $\bracevert t\bracevert\geq\sqrt{M}$ or $\sqrt{M}\leq t\leq+\infty$ (the case $-\infty\leq t\leq -\sqrt{M}$  corresponds to the time reversed solution). The scale factor, in the conformal frame, is then found to be

\begin{equation}
a^\pm=\frac{\sqrt{t^2-M}}{\sqrt{\phi_0}}\exp[\pm\frac{1}{2}\frac{\sqrt{N}}{M}\frac{t}{\sqrt{t^2-M}}],
\end{equation}
while the curvature scalar:

\begin{equation}
R^\pm=\frac{3}{2}N\phi_0 \frac{\exp[\pm\frac{\sqrt{N}}{M}\frac{t}{\sqrt{t^2-M}}]}{(t^2-M)^3}.
\end{equation}

The relationship between the proper time $t$ measured in the untransformed frame and its conformal (for the '+' branch of the solution that is the case of interest) is given by

\begin{equation}
\tau^+=-\frac{t}{\sqrt{M\phi_0}}\exp[\frac{\sqrt{N}}{2M}\frac{t}{\sqrt{t^2-M}}].
\end{equation}

A careful analysis of eq.(4.17) shows that $a^+$ is a minimum at some time that is a root of the algebraic equation $t^4-Mt^2-\frac{N}{4}=0$. The curvature singularity occurring in the Einstein's formulation of general relativity at $t=\sqrt{M}$, is removed in the conformal frame, where $R^+$ is bounded and well behaved for all times in the range $\sqrt{M}\leq t \leq +\infty$ ($-\infty\leq\tau\leq +\infty$).

For a dust-filled universe ($\gamma=1$) the untransformed scale factor can be given the form:

\begin{equation}
\hat a=\frac{4M}{\eta^2-4},
\end{equation}
where the time variable $\eta$ has been introduced through $dt=\frac{\hat a^2}{M}d\eta$ and is constrained to the range $2\leq\eta\leq +\infty$ (the case $-\infty\leq\eta\leq -2$ corresponds to the time reversed solution). In the conformal formulation of GR we have that

\begin{equation}
a^\pm(\eta)=\frac{4M}{\sqrt{\phi_0}}\frac{\exp[\mp\frac{\sqrt{N}}{24M^2}\eta(\eta^2-12)]}{\eta^2-4},
\end{equation}
and the relationship between the proper time $\tau$ and $\eta$ is given by the following expression:

\begin{equation}
\tau^\pm=\frac{16M}{\sqrt{\phi_0}}\int d\eta\frac{\exp[\mp\frac{\sqrt{N}}{24M^2}\eta(\eta^2-12)]}{(\eta^2-4)^2}.
\end{equation}

The curvature singularity occurring in the canonical Einstein's GR at time $\eta=2$ is removed again in the conformal representation of the theory. The '+' branch scale factor $a^+$ is a minimum at some $\eta_*$ that is a root of the algebraic equation $\eta^4-8\eta^2+\frac{16M^2}{\sqrt{N}}\eta+16=0$.

Finally we shall remark the fact that the cosmological singularity is removed, in conformal GR, only for a given range of the parameter $\alpha$ (or $\omega$). It can be taken just as a restriction on the values this parameter can take. A physical consideration why we chose the '+' branch (the non-singular branch) instead of the '-' branch is based on the following analysis. We shall note that in the conformal formulation of GR, $e^{-\hat\phi}$ plays the role of an effective gravitational constant $G$. For the '-' branch $G$ runs from zero to an infinite value, i.e. gravity becomes stronger as the universe evolves and, in the infinite future it dominates over the other interactions, that is in contradiction with the usual picture. On the contrary, for the '+' branch, $G$ runs from an infinite value to zero and hence gravitational effects are weakened as the universe evolves, as required.

\section{Discussion}

In this section, we would like to present some reflections originated by the results we have obtained in this and in the former papers (reference \cite{iq}). These results tell us that, flat and open universes, are free of the cosmological singularity in the conformal formulation of general relativity, in the given region of the parameter space ($-\frac{3}{2}\leq\omega\leq -\frac{4}{3}$, $0<\gamma<2$). Both canonical formulation with the cosmological singularity and the conformal one without them, are just two different representations of the same theory: general relativity. Respecting experimental observations none of these pictures is preferred over the other. However a question is to be raised. Is the cosmological singularity a spurious object (an artifact) of general relativity due to a wrong choice of the representation of this theory?. While trying to answer this fundamental question we should be very careful since experimental observations can not help us. In fact, the untransformed FRW universe with the cosmological singularity and its conformal, FRW, wormhole universe, are related through the conformal transformation (1.1) with $\Omega^2=\phi$, and experimental measurements are insensible to this transformation.

In section {\bf II} we have shown that matter particles non-minimally coupled to the metric $\bf g$, follow the geodesics of Weyl geometry. The metric $\bf g$, together with the scalar field $\phi$, define metrical relations on Weyl spacetimes. Hence, if we follow the line of reasoning, leading to the identification of the metric ${\bf\hat g}$ as the physical metric in canonical GR, we then reach to the conclusion that, $\bf g$ is the physical metric in the conformal representation of general relativity. On the other hand, in section {\bf III}, we showed that conformal general relativity, linked with Weyl geometry, provides a consistent formulation of the laws of gravity. In fact, the effective Lagrangian of this theory (eq.(2.3) or (3.14)), is invariant under the one-parameter group of transformations of the units of length, time and mass studied in that section. Unlike this, canonical general relativity, is not such a consistent theory of spacetime. This means that the singularity-free character of FRW (flat and open) spacetimes, in conformal GR, is relevant enough: for flat and open FRW spacetimes, the answer to the question raised at the beginning of this section, should be positive. The cosmological singularity is an artifact of general relativity, due to the wrong choice of the formulation of this theory.

As it is understood at present, the canonical formulation of general relativity is a classical theory of spacetime. Hence this formulation of GR can not take account of the global singularity at the beginning of time. It is expected that, when a final quantum gravity theory will be worked out, the cosmological singularity will be removed. Unlike this, as we have shown for flat (\cite{iq}) and open universes (present paper), the conformal formulation of general relativity can take account of the full (singularity-free) evolution of the universe in spacetime without resorting to quantum considerations. It is true, however, if the throat radius of the cosmological wormhole, in the frame of conformal GR, is much greater than the Planck length. Since the throat radius depends on the integration constants $N$, $M$, and $\hat\phi_0$ this means that appropriate initial conditions should be given.

\begin{center}
{\bf ACKNOWLEDGMENT}
\end{center}

We acknowledge many colleagues for their interest in the ideas developed in this paper. Their criticism, together with the referee's criticism, contributed to the improvement of the present version of the paper. We, also, thank MES of Cuba by financial support.


\begin{thebibliography}{99}

\bibitem{iq} I. Quiros, gr-qc/9905071 (accepted for publication in Phys. Rev. D); I. Quiros, R. Bonal and R. Cardenas, gr-qc/9908075 (accepted for publication in Phys. Rev. D).
\bibitem{fgn} V. Faraoni, E. Gunzig and P. Nardone, IUCAA 24/98; gr-qc/9811047 (to appear in 'Fundamentals of Cosmic 
Physics').
\bibitem{bdk} C. Brans and R. H. Dicke, Phys. Rev. \textbf{124}, 925(1961).
\bibitem{haw} S. W. Hawking and G. F. R. Ellis, The large scale structure of space-time, 84(Cambridge University Press, Cambridge, 1973).
\bibitem{far} V. Faraoni, Phys. Lett. A \textbf{245}, 26(1998).
\bibitem{dk} R. H. Dicke, Phys. Rev. \textbf{125}, 2163(1962).
\bibitem{gsw} M. B. Green, J. H. Schwarz, E. Witten, Superstring theory, Vol. \textbf{1} (Cambridge University Press, Cambridge, 1987).
\bibitem{vh} M. Visser and D. Hochberg, gr-qc/9710001.

\end{thebibliography}
\end{document}